Broadband tunable mid-infrared absorber based on conductive strip-like meta-atom elements

Henrik Parsamyan[a,b], Hovhannes Haroyan[a], and Khachatur Nerkararyan[a,*]

[a] *Department of Microwave Physics, Yerevan State University, A. Manoogian 1, Yerevan, 0025, Armenia*
[b] *Center for Nanoscience and Technologies, Institute of Chemical Physics, NAS RA, P. Sevak 5/2, Yerevan, 0014, Armenia*

*Corresponding author:

Address: Republic of Armenia, Yerevan, 0025, 1 Alex Manoogian
E-mail address: *knerkar@ysu.am* (Kh. Nerkararyan)

**Abstract**

A metamaterial composed of thin metallic strips as an efficient broadband absorber in the mid-infrared spectrum is investigated. Here the matching between dielectric and geometrical properties of the individual elements is critical to ensure high absorption. Detailed theoretical analysis based on the electric dipole approximation is performed to characterize the absorption and scattering properties of the individual elements of the unit cell and the results are used to design the metamaterial composed of such configurations. The absorption cross-section of a strip-shaped element can exceed its longitudinal geometrical cross-section area by 17 times. It was shown that the absorptance of the Ni-based metal-insulator-metal (MIM) metamaterial structure can exceed 90 % in ~8.2 – 18 μm spectrum. The broadband absorption is associated with the excitation of the low-$Q$-factor dipole modes of the strips. Such an absorber demonstrates good polarization and incident angle tolerance for transverse-electric (TE) waves. The absorption spectrum can be tuned by varying individual element parameters.

1. **Introduction**

Interaction of electromagnetic radiation with subwavelength metal particles and controlling for optical processes such as absorption and scattering is of great interest for both fundamental science and practical applications[1–3]. Particularly, the realization of the high absorption with a simultaneously low scattering in systems consisting of metal particles is important for energy harvesting and conversion[4,5], optoelectronics[6], sensing[7] and investigations of light-matter interactions[8]. Nanoparticles with various shapes, such as spheres, cubes, rods, discs etc. made of noble metals were shown to be powerful tools for label-free biosensing, photodetection, electromagnetic wave energy confinement, optical near-field and absorption enhancement[8–12]. Such characteristics are commonly achieved by exploiting the phenomenon of the localized surface plasmon resonance (LSPR) conditioned by particle shape, material and environment[13]. The LSPR of metal nanoparticles leading to a strong absorption usually lies in the visible and near-infrared spectrum due to high charge carrier concentrations (~$10^{22}$-$10^{23}$ cm$^{-3}$)[8].

Shifting the absorption resonance of individual subwavelength particles to the long-wavelength spectrum (e.g., mid- and long-wave infrared) is not straightforward due to rapid increases in the imaginary part and the absolute value of the real part of the dielectric permittivities of common plasmonic metals[14]. Thus, surface plasmons practically do not appear at longer wavelengths and a strong scattering dominating over the absorption is observed for noble metal particles. However, the plasmonic resonances of doped semiconductor or metal oxide nanocrystals can be tuned over a wide spectral range from visible to far-infrared by choosing the shape of a crystal, dopant type or varying the doping concentration thereby achieving strong absorption at longer wavelengths[12,15,16]. In the review [17] Kriegel et al. presented recent progress in plasmonics of doped semiconductor nanocrystals. In particular, absorption resonances of indium, iron or gallium doped metal oxide nanocrystals, for instance, oxides of zinc, tin, cadmium, molybdenum and others can appear within the short or mid-wave infrared range. However, the absorption spectrum due to LSPR is usually narrowband. Another approach to shifting the absorption resonance towards longer infrared wavelengths with simultaneously negligible scattering is using core-shell configurations[2,14,18]. The absorption then can be controlled by choosing core and shell materials, as well as by varying geometrical parameters such as core radius or shell thickness. In [14] the increased broadband absorption is associated with the configuration resonance of core-shell metal-dielectric spherical and cylindrical systems, where tunable wave-impedance matching of the particles with free-space is ensured by the appropriate choice of the nanometer-thin shell metal and thickness. Furthermore, subwavelength configurations with strong absorption can also be used as unit elements of metasurfaces and metamaterials. Electromagnetic responses of subwavelength configurations randomly or periodically distributed on a substrate mainly define the light interaction with complex systems composed of those configurations[13,19,20]. The emergence of various physical phenomena in metasurfaces are thus conditioned by their structural features (such as periodicity, substrate etc.), as well as by characteristics of the unit-cell element (shape, material and geometrical parameters). As a rule, the unit-cell configurations are chosen to control Joule losses of the system. For instance, one needs to minimize Joule losses to control the intensity and polarization of the reflected and transmitted waves. On the other hand, maximized Joule losses lead to the perfect absorption characteristics of a metasurface. Obtaining reasonable and predictable results while investigating such complex systems requires thorough analysis. Hence, conceptual-based guidelines supported by the theoretical analysis are required to generalize design steps and improve absorption characteristics and for the desired operation of metasurfaces carefully designed unit elements are critical[20–22]. These ideas were widely employed to design efficient near-perfect single-, multi- and broad-band metamaterial absorbers covering the entire electromagnetic spectrum from the visible and infrared to the THz and microwaves[23–27].

In this article, we theoretically and numerically investigate strong mid-wave infrared absorption properties of a subwavelength configuration with an elongated elliptical shape. Theoretical analysis based on electric-dipole quasistatic approximation is carried out to derive resonant conditions for the absorption resonance of an individual particle. The analysis reveals that enhanced absorption dominating over the scattering is ensured for particles made of metals whose absolute values of the real and imaginary parts of dielectric permittivities are of the same order. Obtained analytical expressions will allow one to easily identify the resonant dimensions for the given metal. Finite element method-based simulations are conducted for a rectangular strip as the equivalent and more practical configuration for an elongated oblate ellipsoid, and absorption and scattering characteristics were compared. We show that the absorption cross-sections (ACSs) of a rectangular strip made of nickel (Ni) can exceed its longitudinal geometrical cross-section by about 17. In contrast, the scattering cross-section (SCS) is negligibly small. Increased absorption of such a particle is due to the configuration resonance when a special combination of material properties and geometrical parameters leads to the impedance matching between the configuration and the surrounding space. Using the results for an individual strip, a metamaterial absorber (MMA) composed of a bottom reflector, dielectric spacer and orderly arranged Ni strip-like configurations was designed, whose absorptance exceeds 90% in the broad wavelength range from 8.2 μm to 18 μm. The absorption bandwidth (BW) of ~ 9.8 μm corresponds to the relative bandwidth (RBW) of ~ 75%. The absorption spectrum of the MMA can be effectively varied by changing the length of strip-like elements and metals used. The suggested absorber can find practical applications for energy harvesting and radiative cooling.

## 2. Theory

Consider a prolate ellipsoid with semi-axes such that $a_z \gg a_y > a_x$ and complex dielectric permittivity $\varepsilon_1$ embedded in a lossless dielectric medium $\varepsilon_2$. The ellipsoid is illuminated by a plane wave polarized along its semi-major axis $a_z$. Within the limits of the quasi-static approximation (i.e. $a_z \ll \lambda$, where $\lambda$ is the incident wavelength in the surrounding medium) the dipole moment of a particle is determined by[28,29]:

$$p = \varepsilon_2 \alpha E, \qquad (1)$$

where $\alpha = \varepsilon_0 V \chi$ is the electric diploe polarizability and $V = 4\pi a_x a_y a_z/3$ is the volume of the particle, $\varepsilon_0$ is the vacuum dielectric constant and

$$\chi = \frac{\varepsilon_1 - \varepsilon_2}{\varepsilon_2 + (\varepsilon_1 - \varepsilon_2)n^{(z)}}. \tag{2}$$

Here $n^{(z)}$, which depends only on the shape of the particle[28,30], is the so-called depolarization factor and defined as follows:

$$n^{(z)} = \frac{a_x a_y a_z}{2} \int_0^\infty \frac{ds}{(s + a_z^2)\sqrt{(s + a_z^2)(s + a_y^2)(s + a_x^2)}}. \tag{3}$$

Simplifying the integral leads to the following expression for the depolarization factor of the prolate ellipsoid:

$$n^{(z)} = \frac{\eta\gamma}{(1-\eta^2)\sqrt{1-\gamma^2}}\left[-E(\phi,\kappa) + F(\phi,\kappa)\right], \tag{4}$$

where $\eta = a_y/a_z$, $\gamma = a_x/a_z$, $\kappa = \sqrt{(1-\eta^2)/(1-\gamma^2)}$, and $0 \leq \phi = \cos^{-1}\gamma \leq \pi/2$. $E(\phi,\kappa)$ and $F(\phi,\kappa)$ are Legendre's incomplete elliptic integrals of the first and second kind, respectively[28,30]. Taking into account that $a_x \ll a_z$, for the prolate ellipsoid $E(\phi,\kappa) = \sin\phi \approx 1$ and using the following asymptotic formula for the elliptical integral $F(\phi,\kappa)$[31]:

$$F(\phi,\kappa) \approx \ln\left(\frac{4a_z}{a_y}\right), \tag{5}$$

one obtains the simplified expression of the depolarization factor:

$$n^{(z)} = \eta\gamma \ln\left(\frac{4a_z}{ea_y}\right) \approx \eta\gamma \ln\left(\frac{1.5}{\eta}\right) \tag{6}$$

The polarizability then can be rewritten as follows:

$$\alpha \approx \frac{4\pi\varepsilon_0}{3} \cdot \frac{a_z^3}{\ln(1.5/\eta)} \cdot \frac{(\varepsilon_1 - \varepsilon_2)\eta\gamma\ln(1.5/\eta)}{\varepsilon_2 + (\varepsilon_1 - \varepsilon_2)\eta\gamma\ln(1.5/\eta)}. \tag{7}$$

The absorption and scattering cross-sections of a particle are given by the following expressions[29]:

$$\sigma_{abs} = \frac{k}{\varepsilon_0}\Im m(\alpha), \tag{8}$$

$$\sigma_{scat} = \frac{k^4}{6\pi\varepsilon_0^2}|\alpha|^2. \tag{9}$$

Substituting expressions (6) and (7) into (8), the absorption cross-section then can be written as:

$$\sigma_{abs} \approx \frac{8\pi^2}{\lambda} \frac{a_z^3}{3\ln(1.5/\eta)} \Im m\left(\frac{(\varepsilon_1 - \varepsilon_2)n^{(z)}}{\varepsilon_2 + (\varepsilon_1 - \varepsilon_2)n^{(z)}}\right), \tag{10}$$

Henceforth we assume that the ellipsoid is a metal with the frequency-dependent complex dielectric permittivity $\varepsilon_1(\omega) = \varepsilon_{1r}(\omega) + i\varepsilon_{1i}(\omega)$, where $\varepsilon_{1r}(\omega) < 0$, $\omega = 2\pi c/\lambda$ and $\lambda$ is the operating wavelength. At a given wavelength of the incident field and corresponding dielectric constants of the particle and surrounding, the maximum value of $\Im m(\alpha)$ is achieved when the condition for particle's depolarization factor and material properties is satisfied:

$$n^{(z)} = \frac{|\varepsilon_{1r}|\varepsilon_2}{|\varepsilon_1|^2} \ll 1 \tag{11}$$

For the particular case of the elongated oblate ellipsoid when $a_x = a_y < a_z$, the particle has a form of a prolate spheroid. With all notations used above the expression (6) of the depolarization factor is simplified to:

$$n^{(z)} \approx \eta^2 \ln\left(\frac{1.5}{\eta}\right) \tag{12}$$

Now let's write the condition (11) of the maximum absorption for the ellipsoid in the explicit form:

$$\frac{a_x a_y}{a_z^2} \ln \frac{1.5 a_z}{a_y} = \frac{|\varepsilon_{1r}|\varepsilon_2}{|\varepsilon_1|^2}. \tag{13}$$

One sees that the maximum absorption for a particle is achieved when the certain condition between geometrical parameters and material properties of a particle, as well as surrounding, is satisfied. Thus, depending on the dielectric permittivity of metal and host medium chosen [right-hand side of Eq. (13)], the maximum absorption can be ensured by varying the sizes of a particle so that condition (13) is satisfied.

To quantitatively analyze the absorption and scattering properties of the investigated configurations, we use absorption and scattering efficiency factors defined as follow[14]:

$$Q_{abs} = \frac{\sigma_{abs}}{S} \text{ and } Q_{scat} = \frac{\sigma_{scat}}{S}, \tag{14}$$

where $S = \pi a_y a_z$ is the longitudinal geometrical cross-section area of a particle.

Condition (13) of the absorption maximum of an individual ellipsoid involves several parameters. However, detailed analysis reveals that satisfying the required conditions is not straightforward. The main requirement is that values of the real and imaginary parts of the dielectric permittivity of a metal must be of the same order, and it is desirable that $\varepsilon_{1i} < |\varepsilon_{1r}|$. Moreover, lateral sizes of an ellipse (and those of an equivalent strip, which will be explained next) as a rule need to be smaller than the height by about two orders. Finally, the absorption efficiency must essentially exceed the scattering one:

$$Q_{abs} \gg Q_{scat}, \tag{15}$$

$$Q_{abs} \sim \frac{8\pi a_x |\varepsilon_1|^2}{3\lambda \varepsilon_2 \varepsilon_{1i}} \text{ and } Q_{scat} \sim \frac{(2\pi)^4 a_x^2 a_y a_z |\varepsilon_1|^2}{\lambda^4}. \tag{16}$$

This takes place when the height of a particle is much smaller than the incident wavelength. In the infrared region of the spectrum, a thin and rather long Ni strip satisfies these requirements (the real and imaginary parts of the dielectric constant are of the same order), which was utilized for further numerical analysis. It is noteworthy that the proposed theory is quite universal and can be applied up to the microwave region [20].

## 3. Results and discussion

The theoretical analysis in the frameworks of the electric-dipole approximation was performed for an ellipsoidal particle as it represents a more general case of the configuration. Once the material and geometrical parameters of a configuration are given, derived theoretical relations allow one to easily evaluate the absorption spectrum. Moreover, the absorption resonance can be shifted to the spectrum of interest by varying the geometrical and electrodynamic parameters of a given configuration. However, the validity of the quasi-static approximation is limited by the condition $\lambda \gg a_x, a_y, a_z$ between the incident wavelength and sizes of a particle. As long as the relation $a_z \gg a_y > a_x$ for the ellipsoid dimensions is maintained, the elongated oblate ellipsoidal particle can be approximated to an equivalent rectangular strip which is more practical configuration. Hence, complete parametric analysis of the absorption and scattering properties of the studied configuration was carried out through full-wave numerical simulations based on the finite element method. Theoretical and simulated spectra of the absorption (black) and scattering (red) efficiency factors of ellipsoid-strip made of Ni[32] are represented by solid and symbol-lines in Fig. 1 (a). The geometrical parameters for ellipsoid-strip are: $a_x = d_x/2$, $a_y = d_y/2$ and $a_z = d_z/2$, where $d_x$, $d_y$ and $d_z$ are the width, depth and height of the strip, correspondingly. The longitudinal cross-sectional area of the strip is $S_{strip} = d_y d_z$. One sees that absorption and scattering spectra of the ellipsoid and its equivalent configuration are in good agreement. The absorption resonance of an ellipsoidal particle with $a_x = 5$ nm, $a_y = 25$ nm, $a_z = 1$ μm lies around 11.6 μm with $Q_{abs}$ being about 14 (simulated $Q_{abs} \sim 17$), whereas $Q_{scat}$ is of the order of $\sim 0.44$. The full width at half maximum (FWHM) of the absorption curve of the elliptical particle are about 6.25 μm, which is of the same order as resonant wavelengths. As mentioned above, high absorption efficiencies can be achieved by making use of metals whose complex dielectric constant components are of the same order [see condition (11)]. The choice of Ni for our investigation is due to such optical properties: at 11.56 μm corresponding to the absorption resonance of a strip, $\varepsilon_{Ni} = -1573.3 + 820.42i$[32].

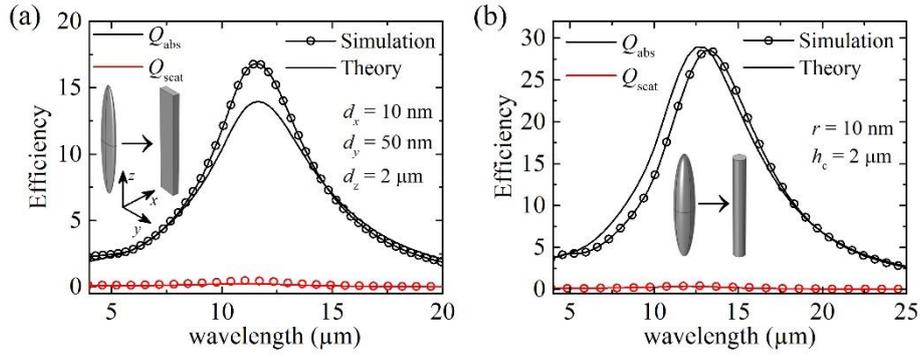

Fig. 1. Theoretical (lines) and simulated (symbols) absorption (black) and scattering (red) efficiency factors of a Ni (a) oblate ellipsoid and rectangular strip with $d_x$ = 10 nm, $d_y$ = 50 nm sizes and (b) prolate spheroid and nanowire with $r$ = 10 nm radius. All configurations have a height of 2 μm. The surrounding medium is air $\varepsilon_2$ = 1.

The numerical calculations also agree well with the theoretical results for a particular case of an elongated oblate ellipsoid when two semi-minor axes are equal corresponding to the case of a rod-like particle. The results for an elongated spheroid and its equivalent rod with sizes $a_x = r$ = 5 nm and $a_z = h_c/2$ = 1 μm, where $r$ is the radius and $h_c$ is the height of the cylinder, are presented in Fig. 1 (b). The longitudinal cross-section area of the cylinder is $S_{cyl} = 2rh_c$. The absorption maximum of the spheroidal Ni particle is around 13.5 μm with the corresponding resonant values of $Q_{abs}$ ~29 and $Q_{scat}$ ~0.4.

Since the enhanced absorption of the strip-like particle is associated with the configuration resonance, it is essential to identify the spectral dependence of the absorption and scattering efficiencies on the geometrical sizes. To do so, we illustrated the simulated absorption and scattering efficiency spectra of the rectangular strip as a function of its depth $d_x$ [Fig. 2 (a) and (b)] and width $d_y$ [Fig. 2 (c) and (d)] by two-dimensional color map plots. $d_x$ is varied from 10 nm to 50 nm and $d_y$ - from 10 nm to 100 nm. In both cases the surrounding medium is air.

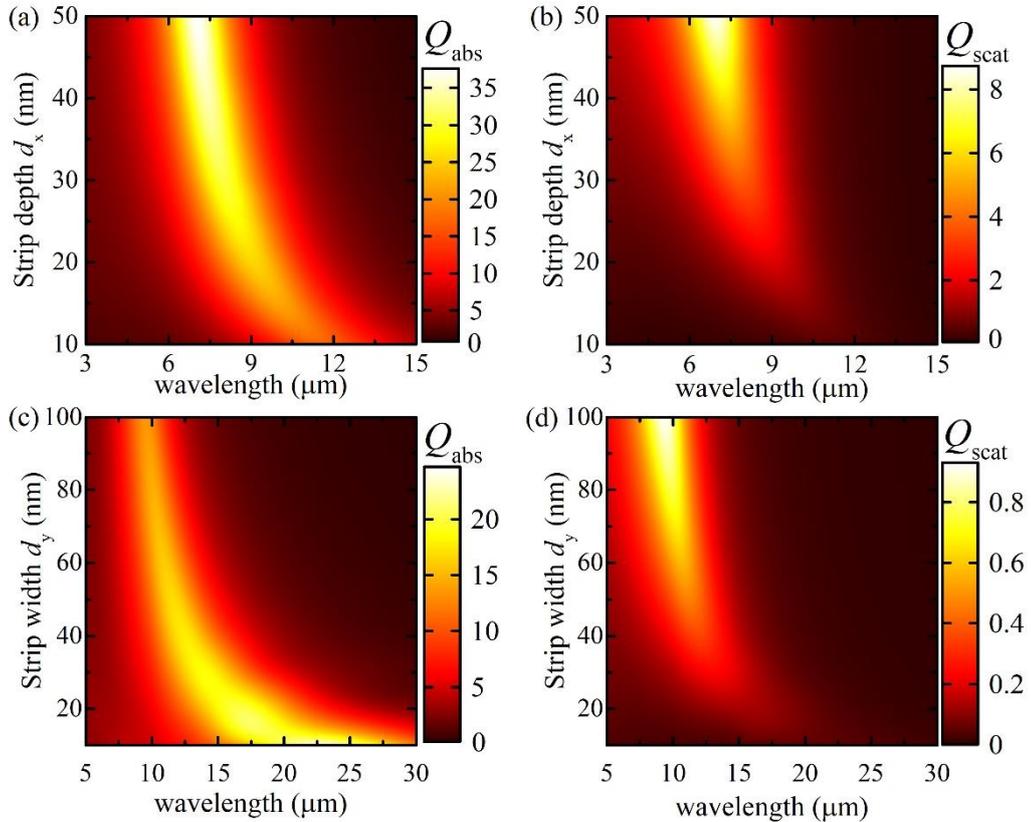

Fig. 2. Two-dimensional colour map plots of absorption and scattering efficiency spectra of a Ni strip as a function of its (a, b) depth $d_x$ ($d_y$ = 50 nm) and (c, d) width $d_y$ ($d_x$ = 10 nm). In both cases strips are 2 μm in height and are surrounded by air.

One sees that an increase in either $d_x$ and $d_y$ leads to a shift of the resonance towards shorter wavelengths and an increase in both efficiency factors. The position of the absorption resonance is determined by the expression (13), where both $\varepsilon_{1r}$ and $\varepsilon_1$ are wavelength-dependent and for an ellipsoid of a given size, the absorption resonance appears at the wavelength for which condition (13) is satisfied. Since the right-hand side of (13) is a monotonically decreasing function depending on the wavelength, an increase in $d_x$ and $d_y$ will lead to a shift of the absorption resonance towards shorter wavelengths. The FWHM is not noticeably changed depending on the $d_x$, but the maximum value of $Q_{abs}$ increases from 17 for $d_x = 10$ nm to about twice for $d_x = 50$ nm. Meanwhile, $Q_{scat}$ becomes essentially large (but still smaller than $Q_{abs}$) at $d_x$ values larger than ~30 nm and reaches 8.2 for $d_x = 50$ nm. On the contrary, both the $Q_{abs}$ and the FWHM decrease as $d_y$ increases, at the same time, changes in $Q_{scat}$ are relatively insignificant. As far as the possibility of infrared absorption at longer wavelengths is concerned, one needs to consider spheroidal (cylindrical) particles with a radius of a few tens of nanometers rather than elliptical (strip-like) ones as follows from Fig. 2 (c). Practical applications of a strip-like configuration may also be limited by the fact that the scattering sharply increases as width of the strip increases [see Fig. 2(b)].

Importantly, the considered configurations can serve as basic elements of an absorber unit cell. To show this, we designed a metamaterial absorber consisting of a Ni reflector, a dielectric spacer ($n_d = 1.5$) and orderly arranged rectangular Ni strips as unit-cell elements. Three-dimensional schematic of the metamaterial absorber and the top-view of the unit cell are illustrated in Fig. 3(a). $p_y$ and $p_z$ are the periodicities of the structure. The unit cell of the absorber is illuminated by a $z$-polarized plane wave propagating along the $x$-axis. Floquet periodic boundary conditions were applied at the $XZ$ and $XY$ planes of the unit cell. Throughout analysis optimized dimensions for the unit cell $p_y = 400$ nm and $p_z = 2.7$ μm (The choice of such values will be explained later) and strip $d_x = 10$ nm, $d_y = 50$ nm and $d_z = 2$ μm were used if not otherwise stated. Optimization is conducted aiming to broaden the absorption band and to increase the overall absorption level.

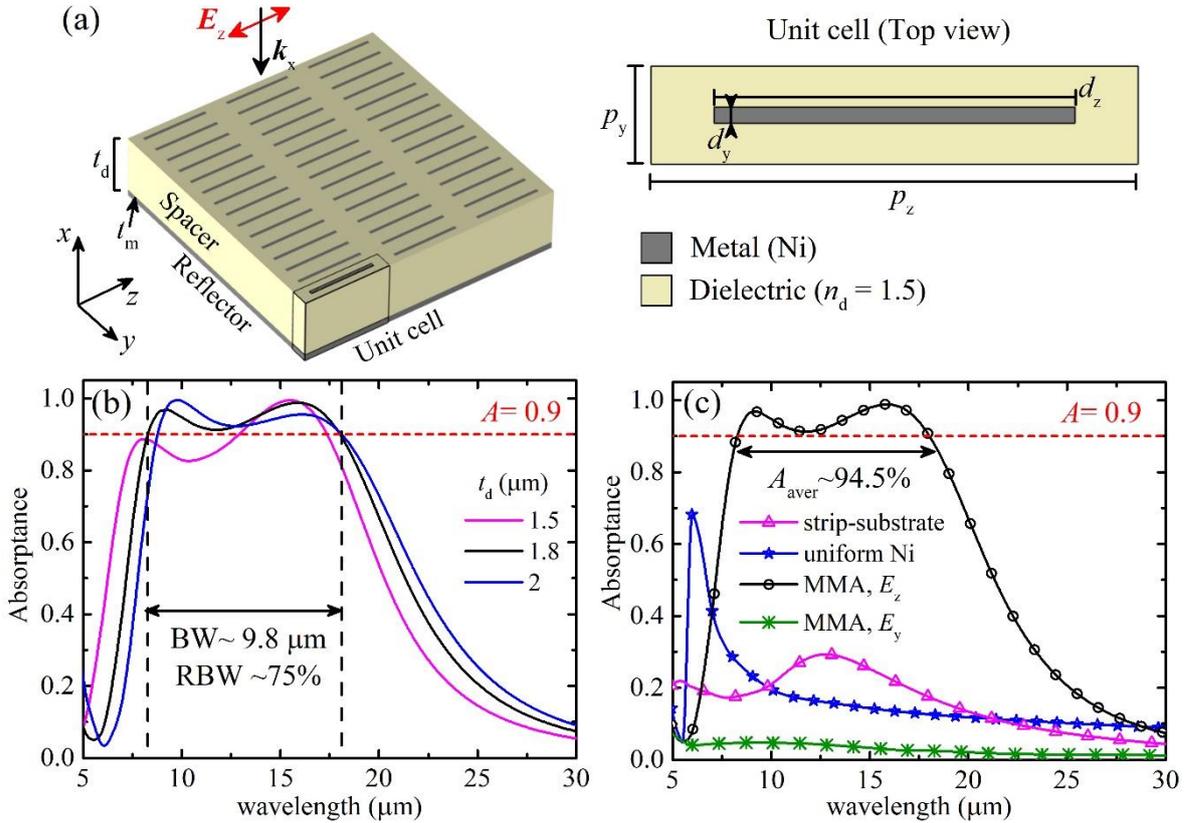

Fig. 3. (a) 3D schematic of the periodic structure composed of a Ni reflector, dielectric spacer with $n_d = 1.5$ and rectangular Ni strips as a unit-cell element; and the top-view of the unit cell. $t_m$ and $t_d$ are the thickness of the reflector and spacer, respectively. (b) The absorption spectrum as a function of the thickness of the dielectric spacer. (c) Comparison of the absorption spectra of a periodic structure composed of a dielectric substrate and Ni strips only (purple-triangle), Ni-dielectric-Ni planar structure (blue-star), and proposed MMA under $z$- and $y$-polarized plane waves (black-ring and green-asterisk, respectively). The geometrical parameters are $t_d = 1.8$ μm, $p_y = 400$ nm, $p_z = 2.7$ μm, $d_x = 10$ nm, $d_y = 50$ nm and $d_z = 2$ μm if otherwise is not stated.

The sizes of the strip are as used previously. Generally, the absorptance $A$ is calculated according to the expression $A(\lambda)=1-R(\lambda)-T(\lambda)$, where $R=|S_{11}|^2$ and $T=|S_{21}|^2$ are the reflectance and transmittance expressed via wavelength-dependent scattering $S$-parameters. Since the 200-nm-thick Ni reflector prevents the transmission through the structure and hence $T = 0$, the absorptance is calculated by the simplified expression $A(\lambda)=1-R(\lambda)$. Fig. 3 (b) plots the spectrum of the absorptance of the structure for the dielectric spacer thicknesses of 1.5 μm (purple), 1.8 μm (black) and 2 μm (blue). It is seen that increasing the spacer thickness results in an increase in the absorptance of the structure so that for the system with $t_d = 1.8$ μm the absorption band over 90 % ($A \geq 0.9$) ranges from about 8.2 up to 18 μm. The absorption bandwidth (BW) in this case is ~ 9.8 μm. The BW relative to the center wavelength $\lambda_c$ of the absorption band defined as RBW = BW/$\lambda_c$ × 100 % is about 75%. The total thickness of the optimized absorber is 2.01 μm, which is around of 0.15$\lambda_c$. Note that the BW of the MMA with $t_d = 2$ μm is slightly smaller.

Absorption performances of the strip-based metamaterial absorber, the periodic structure composed of a dielectric substrate and Ni strip only (without a bottom reflector) and Ni-dielectric-Ni planar structure are compared in Fig. 3 (c). In all cases here the thickness of the dielectric substrate and top Ni layer (uniform or strip-like) is 1.8 μm and 10 nm, respectively. The maximum absorption of strip-dielectric structure is approximately 0.3. The absorption spectrum of the dielectric spacer sandwiched by 200-nm- and 10-nm-thick bottom and top uniform Ni layers has only one resonant peak with $A \sim 0.7$, which is associated with a Fabry-Perot mode of the system. As the meta-atom element of the proposed MMA features highly structural anisotropy, the absorptance is sensitive to the incident field polarization. The absorption spectra of the MMA under parallel ($E_z$)- and perpendicular ($E_y$)-polarized plane waves are shown in Fig. 3(c) by back-rings and green-asterisks, respectively. The absorptance of the MMA $A = 0.1$ in the case of a $y$-polarized plane wave. However, by introducing rectangular Ni strips parallel to the incident field polarization the absorption is essentially increased so that the average absorptance over the high absorption band (above 0.9) is about 94.5 %.

The absorption mechanism of the proposed MMA can be revealed through the electromagnetic field distributions of the structure. Fig. 4 (a) shows the $xz$-plane distributions of the squares of magnitudes of the normalized electric (the first column) and magnetic (the second column) fields of the structure under a $z$-polarized plane wave of a wavelength of 6 μm and 15 μm (the first and second rows). These values correspond to the near-perfect reflection and absorption, respectively (see Fig. 3 (b)). Here three unit-cells in the $z$-direction are shown for illustrative purposes. One sees that at $\lambda = 6$ μm the electromagnetic field is totally reflected from the bottom metallic layer and the strips do not interact with the incident field. The situation is different at $\lambda = 15.8$ μm. Strong absorption of the system is facilitated by a combination of two main processes. Low quality $Q$-factor electric dipole modes excited in the strip and their image due to the bottom metal layer form opposite currents so that a strong magnetic field occurs between the top and bottom layers [see Fig. 4 (b)].

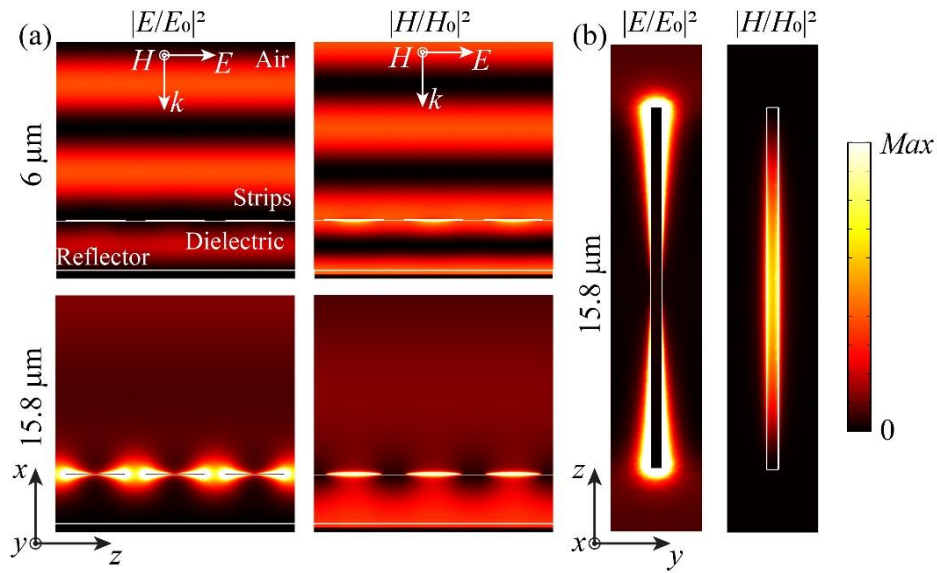

Fig. 4. The distributions of the squares of magnitudes of normalized electric and magnetic fields of the MMA in the (a) incident $zx$-plane at 6 μm and 15.8 μm; (b) The same distributions in the $yz$-plane under the incident field wavelength of 15.8 μm.

Hence, in the far-field, the reflectance can be completely vanished by the destructive interference of these two dipoles resulting from the π phase shift[23]. Near-perfect absorption is thus associated with these localized electric and magnetic dipole resonances, which provide enough time for light energy to be absorbed due to ohmic losses in the metals [Fig 4(b)]. Here also the role of destructive interference due to multiple reflections from the top and bottom layers of the metamaterial should be noted. To ensure this, the reflection coefficient and phase of the metamaterial should be controlled[33,34]. Fig 4(b) shows the same at the wavelength of 15.8 μm in the lateral cross-section of a unit cell ($yz$-plane). Low $Q$-factor of dipole modes of a strip are facilitated by the large value of the loss tangent of Ni in the mid-infrared (at 15.8 μm tan$\delta$ ~ 0.585).

Fig. 5 (a) and (b) show dependences of the absorption spectrum on the structure periodicities $p_y$ and $p_z$, respectively. Dimensions of the unit-cell are optimized to ensure a broad absorption band with absorptance exceeding the $A=0.9$ value. For this, values of periodicities $p_y$ = 400 nm and $p_z$ = 2.7 μm were found to be optimal. It is noteworthy that the efficient broadband absorption is also ensured by utilizing other highly lossy metals such as Cr and Fe. These metals are also characterized by the real and imaginary parts of the dielectric permittivity of the same order in magnitude so that loss tangents are rather large.

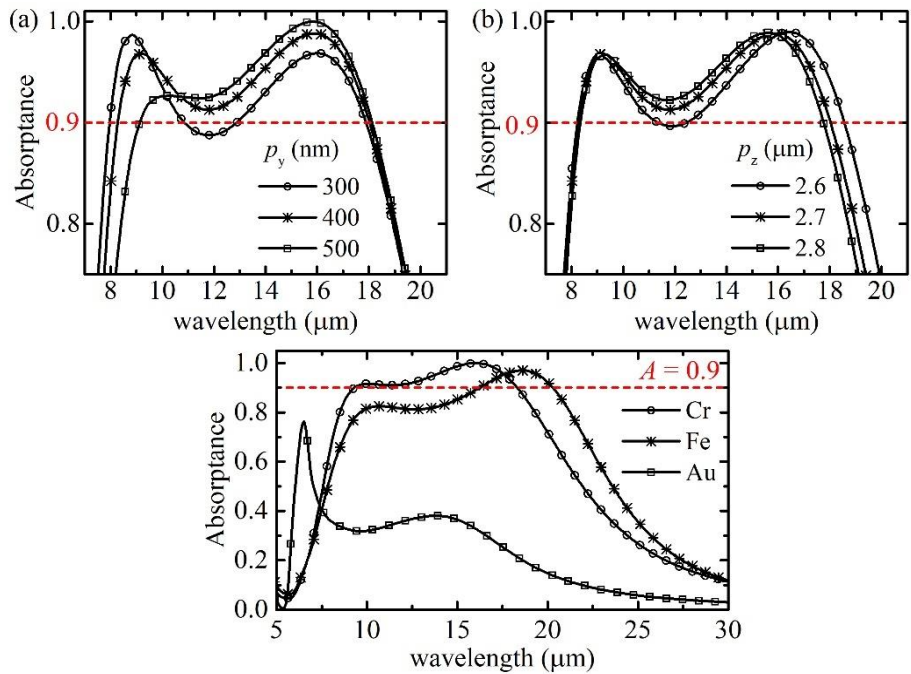

Fig. 5. The absorption spectrum at different values of unit-cell periodicities (a) $p_y$ (300, 400 and 500 nm) and (b) $p_z$ (2.6, 2.7 and 2.8 μm). (c) The absorption spectra for Au-, Cr- and Fe-based absorber. Materials for both the bottom reflector and strips were the same during the calculations. The geometrical parameters are as indicated in caption of Fig. 3.

For instance, the above-0.9 absorption band of Cr-based structure ranges from 9.2 up to 18.35 μm and the BW = 9.15 μm. Meanwhile, the Fe-based structure demonstrates efficient absorption exceeding 0.8 in the 9.6-21.24 μm spectrum (BW = 11.65 μm). In contrast, the Ni-based structure shows above-0.8 absorption with a BW of 11.1 μm. On the contrary, employing a conventional plasmonic metal - Au result in only ~ 30 % of the incident radiation to be absorbed in the range of ~5.8-16.5 μm.

Fig. 6 (a) and (b) depict the spectrum of the absorptance of the Ni-based absorber as a function of the polarization ($\varphi$) and incidence ($\alpha$) angle, respectively, and their corresponding schematics. One sees that the electromagnetic waves polarized along the long axis of the strip ($\varphi = 0°$) are efficiently absorbed in the broad spectrum. However, as the polarization angle of the incident field increases, the reflection also increases and in the extreme case when the polarization of the incident light becomes perpendicular to the long axis of the strip ($\varphi = 90°$), waves are completely reflected. The absorber also shows good incidence-angle tolerance for TE-polarized electromagnetic waves [see Fig. 6 (b)]. A broad efficient absorption band is observed at incident angles up to $\alpha = 40°$, and a further increase in $\alpha$ leads to a distortion of the absorption band. However, two nearly-perfect absorption peaks are still observed at incident angles above $\alpha = 40°$.

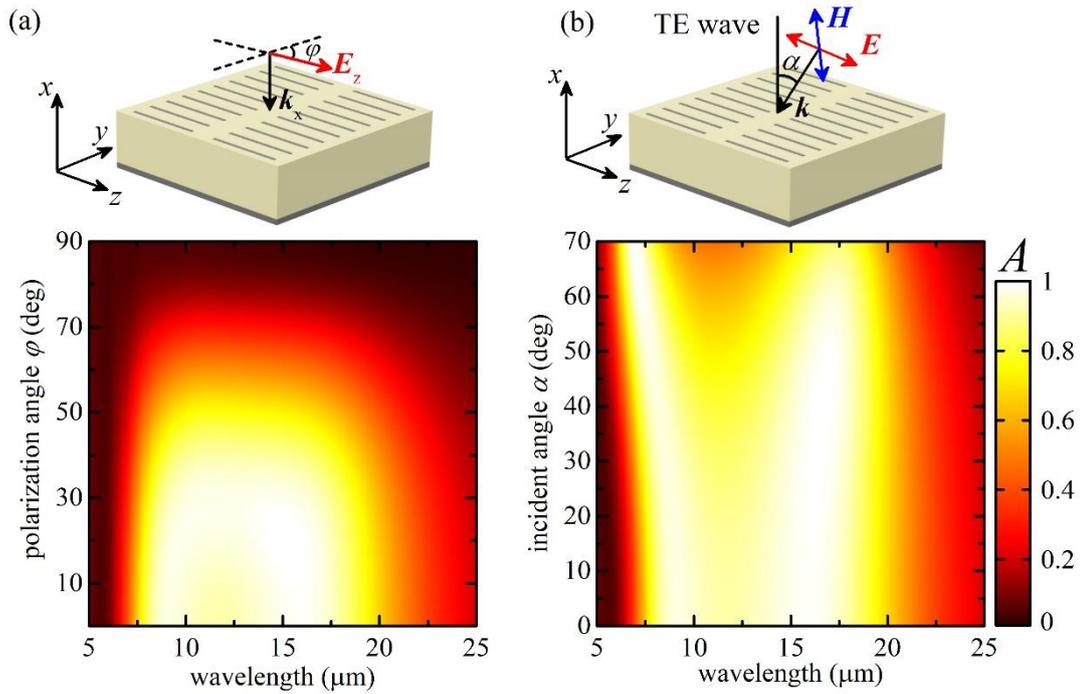

Fig. 6. Two-dimensional colour map plots of absorptance spectrum of the structure as a function of the (a) polarization angle $\varphi$ defined as the angle between the strip long axis and incident electric field polarization, and (b) incidence angle $\alpha$ in the case of TE-polarized waves with the corresponding schematic illustrations.

Although such an absorber is polarization-sensitive, another parallel layer with strips oriented perpendicular to the first layer elements can be added to ensure polarization insensitivity. It should be emphasized that in the case of polarization of the incident field perpendicular to the configuration axis neither absorption nor reflection is produced from the top layer of periodic strips, and electromagnetic waves are completely reflected back from the bottom metal layer.

In Table 1 performance of the investigated absorber in our work with other metamaterial absorbers operating in the mid-infrared spectrum. Since after some design modifications the suggested absorber can be made polarization-insensitive, e.g., by adding another layer of strips perpendicular to the first one, both polarization-dependent and -independent absorbers are considered.

**Table 1**
Comparison of performance of the suggested absorber with other metamaterial absorbers operating in the same spectrum

| Ref. | Polarization sensitivity | 90% absorption band (BW) in μm | RBW (%) | Unit cell sizes (μm³) | Materials | Layers |
|---|---|---|---|---|---|---|
| [35] | No | 8 – 14 (6) | 55 | 7.4 × 7.4 × 1.922 | Ti, Ge | 5 |
| [36] | No | 8 – 13 (5) | 48 | 3.1 × 3.1 × 2.105 | Ti, $SiO_2$ | 3 |
| [37] | Yes | 8 – 14 (6) | 55 | 1.6 × 1.6 × 0.91 | Ti, Si | 3 |
| [38] | No | 8 – 14 (6) | 55 | 1.6 × 1.6 × 0.82 | Ti, Ge, $Si_3N_4$ | 4 |
| [39] | No | 7.7 – 12.2 (4.5) | 45 | 6.76 × 6.76 × 0.79 | ZnS, Au | 3 |
|  |  | 5.2 – 13.7 (8.5) | 90 | 9.2 × 9.2 × 1.56 | $Al_2O_3$, ZnS, Au | 5 |
| Our work | Yes | 8.2 – 18 (9.8) | 75 | 0.4 × 2.7 × 2.01 | Ni, dielectric ($n_d$ = 1.5) | 3 |

Other criteria of comparison are above-90% absorption bandwidth, relative bandwidth, unit cell sizes (periodicities and the total thickness), materials used and the number of layers (including the bottom reflector if used). One sees that the performance and geometrical sizes of the suggested metamaterial absorber are comparable with those lately reported in the literature. A relatively broad and highly efficient absorption band here is achieved by only three structural layers and small unit-cell dimensions compared with the center wavelength of the absorption band ($\lambda_c$ = 13.1 μm). The structure of the absorber is very simple and the fabrication process is rather straightforward. The first step of the fabrication is covering the dielectric substate by a 200-nm-thick metal layer making use of conventional deposition methods. Ni strips on the top of the MMA then can be fabricated by either mask lithography or laser-direct-write process[23,24]. The applications of metamaterial absorbers range from sensing, energy harvesting, conversion, thermal emission, and cloaking to optoelectronics. The operation range of the proposed MMA is 8.2-18 μm, which includes the infrared transparency window of the Earth's atmosphere (8–13 μm). This window is used for radiative cooling intended to radiate the heat caused by solar radiation into the universe thereby achieving passive, energy-free cooling. One of the main advantages of the suggested MMA is that the operating spectrum can be effectively tuned to the desired range by varying the geometrical sizes of the strip and unit cell. So, employing such MMA thermal emitters in various electromagnetic spectra can be designed. The simple meta-atom element chosen allows one to derive analytical expressions to reveal the most favourable conditions for the absorption resonance depending on the geometrical parameters at given material properties. Importantly, these conditions are universal and can be used to design an absorber operating from microwave to optical frequencies. Moreover, numerical simulations demonstrate that the performance of an absorber based on such simple meta-atom elements is comparable with that of other MMAs reported in the literature, as presented in Table 1. Unit cell sizes of the proposed MMA are much smaller than the center wavelength of the absorption band. So, the sizes of the unit-cell relative to the $\lambda_c$ are $p_x \times p_y \times t_{total}$ = 0.03 × 0.2× 0.15.

## 4. Conclusions

Thus, efficient broadband absorption of a metamaterial can be ensured not only via the proper mutual arrangement of the unit-cell elements with a stripline shape but also through the appropriate choice of the geometrical and dielectric parameters of the individual elements. Step-by-step guidelines in the frame of electrostatic approximation for maximizing the absorption cross-sections of an individual ellipsoidal particle caused by the configuration resonance are established. Efficient broadband metamaterial absorber composed of such configurations as unit-cell elements was designed. Controllable and high absorption by such metamaterials in the mid-infrared spectrum can be obtained by making use of highly lossy metals whose real and imaginary parts of the dielectric permittivity are of the same order of magnitude. Particularly, the absorptance of the metamaterial composed of bottom Ni reflector, dielectric spacer and orderly arranged rectangular Ni strips exceeds 90 % in a broad spectrum from ~8.2 up to 18 μm. This is also facilitated by the appropriate choice of geometrical parameters of the elements. Remarkably the absorption spectrum can be tuned by the corresponding selection of parameters of the configuration units.

### Acknowledgement

This work was supported by scientific research grants through the SC of MESCS of Armenia (20DP-1C05, 20APP-1C009 and 21AG-1C061).

### Data availability

The data that support the findings of this study are available from the corresponding authors upon reasonable request.

### References

[1]  R.W. Lord, J. Fanghanel, C.F. Holder, I. Dabo, R.E. Schaak, Colloidal Nanoparticles of a Metastable Copper Selenide Phase with Near-Infrared Plasmon Resonance, Chem. Mater. 32 (2020) 10227–10234. https://doi.org/10.1021/acs.chemmater.0c04058.

[2] J. Zhu, L. Meng, G. Weng, J. Li, J. Zhao, Tuning quadruple surface plasmon resonance in gold nanoellipsoid with platinum coating: from ultraviolet to near infrared, Appl. Phys. A 2021 1278. 127 (2021) 1–10. https://doi.org/10.1007/S00339-021-04749-6.

[3] J. Liu, L. Chu, Z. Yao, S. Mao, Z. Zhu, J. Lee, J. Wang, L.A. Belfiore, J. Tang, Fabrication of Au network by low-degree solid state dewetting: Continuous plasmon resonance over visible to infrared region, Acta Mater. 188 (2020) 599–608. https://doi.org/10.1016/j.actamat.2020.02.050.

[4] F. Seyedheydari, K. Conley, T. Ala-Nissila, Near-IR Plasmons in Micro and Nanoparticles with a Semiconductor Core, Photonics. 7 (2020) 10. https://doi.org/10.3390/photonics7010010.

[5] M.A. Baqir, Wide-band and wide-angle, visible- and near-infrared metamaterial-based absorber made of nanoholed tungsten thin film, Opt. Mater. Express. 9 (2019) 2358. https://doi.org/10.1364/OME.9.002358.

[6] S. Kasani, K. Curtin, N. Wu, A review of 2D and 3D plasmonic nanostructure array patterns: fabrication, light management and sensing applications, Nanophotonics. 8 (2019) 2065–2089. https://doi.org/10.1515/NANOPH-2019-0158.

[7] C. Singh, M. Thiele, A. Dathe, S. Thamm, T. Henkel, G. Sumana, W. Fritzsche, A. Csáki, Tri-sodium citrate stabilized gold nanocubes for plasmonic glucose sensing, Mater. Lett. 304 (2021) 130655. https://doi.org/10.1016/J.MATLET.2021.130655.

[8] L. Jauffred, A. Samadi, H. Klingberg, P.M. Bendix, L.B. Oddershede, Plasmonic Heating of Nanostructures, Chem. Rev. 119 (2019) 8087–8130. https://doi.org/10.1021/acs.chemrev.8b00738.

[9] S. Arslanagić, R.W. Ziolkowski, Highly Subwavelength, Superdirective Cylindrical Nanoantenna, Phys. Rev. Lett. 120 (2018) 237401. https://doi.org/10.1103/PhysRevLett.120.237401.

[10] O.A. Balitskii, Recent energy targeted applications of localized surface plasmon resonance semiconductor nanocrystals: a mini-review, Mater. Today Energy. 20 (2021) 100629. https://doi.org/10.1016/j.mtener.2020.100629.

[11] A. Agrawal, I. Kriegel, D.J. Milliron, Shape-Dependent Field Enhancement and Plasmon Resonance of Oxide Nanocrystals, J. Phys. Chem. C. 119 (2015) 6227–6238. https://doi.org/10.1021/acs.jpcc.5b01648.

[12] T. Li, V. Nagal, D.H. Gracias, J.B. Khurgin, Sub-wavelength field enhancement in the mid-IR: photonics versus plasmonics versus phononics, Opt. Lett. 43 (2018) 4465. https://doi.org/10.1364/OL.43.004465.

[13] M. Oh, E.S. Carlson, T.E. Vandervelde, Coupled resonance via localized surface plasmon polaritons in Iridium-based refractory metamaterials, Comput. Mater. Sci. 197 (2021) 110598. https://doi.org/10.1016/j.commatsci.2021.110598.

[14] H.A. Parsamyan, K. V. Nerkararyan, S.I. Bozhevolnyi, Efficient broadband infrared absorbers based on core-shell nanostructures, J. Opt. Soc. Am. B. 36 (2019) 2643. https://doi.org/10.1364/JOSAB.36.002643.

[15] W.T. Hsieh, P.C. Wu, J.B. Khurgin, D.P. Tsai, N. Liu, G. Sun, Comparative Analysis of Metals and Alternative Infrared Plasmonic Materials, ACS Photonics. 5 (2018) 2541–2548. https://doi.org/10.1021/acsphotonics.7b01166.

[16] M.A. Baqir, Conductive metal–oxide-based tunable, wideband, and wide-angle metamaterial absorbers operating in the near-infrared and short-wavelength infrared regions, Appl. Opt. 59 (2020) 10912. https://doi.org/10.1364/AO.411268.

[17] I. Kriegel, F. Scotognella, L. Manna, Plasmonic doped semiconductor nanocrystals: Properties, fabrication, applications and perspectives, Phys. Rep. 674 (2017) 1–52.

https://doi.org/10.1016/j.physrep.2017.01.003.

[18] M. Wan, Y. Li, J. Chen, W. Wu, Z. Chen, Z. Wang, H. Wang, Strong tunable absorption enhancement in graphene using dielectric-metal core-shell resonators, Sci. Rep. 7 (2017) 32. https://doi.org/10.1038/s41598-017-00056-4.

[19] B. Zhao, J. Deng, L. Liang, C. Zuo, Z. Bai, X. Guo, R. Zhang, Lightweight porous Co 3 O 4 and Co/CoO nanofibers with tunable impedance match and configuration-dependent microwave absorption properties, CrystEngComm. 19 (2017) 6095–6106. https://doi.org/10.1039/C7CE01464C.

[20] H. Parsamyan, H. Haroyan, K. Nerkararyan, Broadband microwave absorption based on the configuration resonance of wires, Appl. Phys. A. 126 (2020) 773. https://doi.org/10.1007/s00339-020-03964-x.

[21] A.I. Barreda, J.M. Saiz, F. González, F. Moreno, P. Albella, Recent advances in high refractive index dielectric nanoantennas: Basics and applications, AIP Adv. 9 (2019) 040701. https://doi.org/10.1063/1.5087402.

[22] C. Liang, Z. Yi, X. Chen, Y. Tang, Y. Yi, Z. Zhou, X. Wu, Z. Huang, Y. Yi, G. Zhang, Dual-Band Infrared Perfect Absorber Based on a Ag-Dielectric-Ag Multilayer Films with Nanoring Grooves Arrays, Plasmonics. 15 (2020) 93–100. https://doi.org/10.1007/s11468-019-01018-4.

[23] S. Ogawa, M. Kimata, Metal-Insulator-Metal-Based Plasmonic Metamaterial Absorbers at Visible and Infrared Wavelengths: A Review, Materials (Basel). 11 (2018) 458. https://doi.org/10.3390/ma11030458.

[24] H. Parsamyan, Near-perfect broadband infrared metamaterial absorber utilizing nickel, Appl. Opt. 59 (2020) 7504. https://doi.org/10.1364/AO.398609.

[25] T. Van Huynh, B.S. Tung, B.X. Khuyen, S.T. Ngo, V.D. Lam, N.T. Tung, Controlling the absorption strength in bidirectional terahertz metamaterial absorbers with patterned graphene, Comput. Mater. Sci. 166 (2019) 276–281. https://doi.org/10.1016/j.commatsci.2019.05.011.

[26] R.M.H. Bilal, M.A. Naveed, M.A. Baqir, M.M. Ali, A.A. Rahim, Design of a wideband terahertz metamaterial absorber based on Pythagorean-tree fractal geometry, Opt. Mater. Express. 10 (2020) 3007. https://doi.org/10.1364/OME.409677.

[27] Z. Wang, X. Wang, J. Wang, H. Pang, S. Liu, H. Tian, Independently tunable dual-broadband terahertz absorber based on two-layer graphene metamaterial, Optik (Stuttg). 247 (2021) 167958. https://doi.org/10.1016/j.ijleo.2021.167958.

[28] L.D. Landau, L.P. Pitaevskii, E.M. Lifshitz, Electrodynamics of Continuous Media, 2nd ed., ELSEVIER SCIENCE & TECHNOLOGY, Oxford, United Kingdom, 1984.

[29] R.D. Averitt, S.L. Westcott, N.J. Halas, Linear optical properties of gold nanoshells, J. Opt. Soc. Am. B. 16 (1999) 1824. https://doi.org/10.1364/JOSAB.16.001824.

[30] Carlos E. Solivérez, Electrostatics and Magnetostatics of Polarized Ellipsoidal Bodies: The Depolarization Tensor Method, Free Scientific Information, 2016.

[31] B.C. Carlson, J.L. Gustafson, Asymptotic Expansion of the First Elliptic Integral, SIAM J. Math. Anal. 16 (1985) 1072–1092. https://doi.org/10.1137/0516080.

[32] M.A. Ordal, R.J. Bell, R.W. Alexander, L.L. Long, M.R. Querry, Optical properties of Au, Ni, and Pb at submillimeter wavelengths, Appl. Opt. 26 (1987) 744. https://doi.org/10.1364/AO.26.000744.

[33] M.A. Naveed, R.M.H. Bilal, M.A. Baqir, M.M. Bashir, M.M. Ali, A.A. Rahim, Ultrawideband fractal metamaterial absorber made of nickel operating in the UV to IR spectrum, Opt. Express. 29 (2021) 42911. https://doi.org/10.1364/OE.446423.

[34] W. Pan, T. Shen, Y. Ma, Z. Zhang, H. Yang, X. Wang, X. Zhang, Y. Li, L. Yang, Dual-band and polarization-independent metamaterial terahertz narrowband absorber, Appl. Opt. 60 (2021) 2235. https://doi.org/10.1364/AO.415461.

[35] Y. Luo, D. Meng, Z. Liang, J. Tao, J. Liang, C. Chen, J. Lai, T. Bourouina, Y. Qin, J. Lv, Y. Zhang, Ultra-broadband metamaterial absorber in long wavelength Infrared band based on resonant cavity modes, Opt. Commun. 459 (2020) 124948. https://doi.org/10.1016/j.optcom.2019.124948.

[36] L. Li, H. Chen, Z. Xie, W. Chen, W. Zhang, W. Liu, L. Li, Ultra-broadband metamaterial absorber for infrared transparency window of the atmosphere, Phys. Lett. A. 383 (2019) 126025. https://doi.org/10.1016/j.physleta.2019.126025.

[37] Z. Qin, D. Meng, F. Yang, X. Shi, Z. Liang, H. Xu, D.R. Smith, Y. Liu, Broadband long-wave infrared metamaterial absorber based on single-sized cut-wire resonators, Opt. Express. 29 (2021) 20275. https://doi.org/10.1364/OE.430068.

[38] Y. Zhou, Z. Liang, Z. Qin, X. Shi, D. Meng, L. Zhang, X. Wang, Broadband long wavelength infrared metamaterial absorbers, Results Phys. 19 (2020) 103566. https://doi.org/10.1016/j.rinp.2020.103566.

[39] W. Guo, Y. Liu, T. Han, Ultra-broadband infrared metasurface absorber, Opt. Express. 24 (2016) 20586. https://doi.org/10.1364/OE.24.020586.